\documentclass[twocolumn]{jpsj3}
\usepackage{txfonts}
\usepackage{graphicx}
\usepackage{dcolumn}
\usepackage{bm}

\def\lesssim{\ \raise.3ex\hbox{$<$}\kern-0.8em\lower.7ex\hbox{$\sim$}\ }
\def\gesim{\ \raise.3ex\hbox{$>$}\kern-0.8em\lower.7ex\hbox{$\sim$}\ }

\title{Strong-coupling Properties of a $p$-wave Interacting Fermi Gas on the Viewpoint of Specific Heat at Constant Volume}

\author{Daisuke Inotani\thanks{dinotani@rk.phys.keio.ac.jp}, Pieter van Wyk, Yoji Ohashi}
\inst{Department of Physics, Keio University, 3-14-1 Hiyoshi, Kohoku-ku, Yokohama 223-8522, Japan} 
\abst{We theoretically investigate the specific heat $C_V$ at constant volume in the normal state of a $p$-wave interacting Fermi gas. Including fluctuations in the $p$-wave Cooper channel within the framework of the strong-coupling theory developed by Nozi\`eres and Schmitt-Rink, we clarify how $C_V$ as a function of temperature varies, as one moves from the weak-coupling regime to the strong-coupling limit. In the weak-coupling regime, $C_V$ is shown to be enhanced by $p$-wave pairing fluctuations, near the superfluid phase transition temperature $T_{\rm c}$. Similar enhancement of $C_V(T\simeq T_{\rm c})$ is also obtained in the strong-coupling regime, which, however, reflects that system is close an ideal Bose gas of $p$-wave two-body bound molecules. Using these results, we classify the normal state into (1) the normal Fermi gas regime, (2) the $p$-wave molecular Bose gas regime, and (3) the region between the two, where $p$-wave pairing fluctuations are dominant. Since the current experiments can only access the normal phase of a $p$-wave interacting Fermi gas, our results would be useful for experiments to understand strong-coupling properties of this Fermi system above $T_{\rm c}$.
} 
\begin{document}
\maketitle
%%%%%%%%%%%%%%%%%%%%%%%%%%%%%%%%%%%%%%%%%%%%%%%%%%%%%%%%%%%%%%%%%%%%%%%%%%%%%%%
\par
%%%%%%%%%%%%%%%%%%%%%%%%%%%%%%%%%%%%%%%%%%%%%%%%%%%%%%%%%%%%%%%%%%%%%%%%%%%%%%
\section{Introduction}
Although an ultracold Fermi gas with a tunable $p$-wave pairing interaction has already been realized\cite{Regal,Regal2,Ticknor,Zhang,Schunck,Gunter,Gaebler2,Inaba,Fuchs,Mukaiyama,Maier,Chin}, the study of this Fermi system has not progressed very much, compared to the $s$-wave case\cite{Chin,Zwerger}. One reason is that the $p$-wave pairing interaction causes serious three-body loss\cite{Castin,Gurarie3}, as well as dipolar relaxation\cite{Gaebler2}, leading to very short lifetime of $p$-wave pairs ($=5\sim 20$ ms)\cite{Chevy}. This unwanted effect prevents us from reaching the $p$-wave superfluid phase transition, because this lifetime is much shorter than the typical time scale of condensation growth ($=O(100~{\rm ms})$). Thus, in the current stage of research on a $p$-wave interacting Fermi gas, it is reasonable to start from normal state properties. Of course, the above-mentioned problem also exists in the normal state above the $p$-wave superfluid phase transition temperature $T_{\rm c}$. However, as least, we do not have to wait the ``growth of the system", in contrast to the superfluid case.
\par
In this paper, we theoretically investigate how normal state properties of a $p$-wave interacting Fermi gas vary, as one moves from the weak-coupling regime to the strong-coupling limit. In the $s$-wave case, this problem has extensively been discussed in the context of the so-called BCS (Bardeen-Cooper-Schrieffer)-BEC (Bose-Einstein condensation) crossover phenomenon\cite{Chin,Zwerger,NSR,OhashiGriffin}. In this many-body phenomenon, below $T_{\rm c}$, the character of a Fermi superfluid continuously changes from the weak-coupling BCS-type to the BEC of tightly bound molecules that have already been formed above $T_{\rm c}$, as one passes through the BCS-BEC crossover region. Above $T_{\rm c}$, an atomic Fermi gas continuously changes to a molecular Bose gas with increasing the strength of an $s$-wave interaction. In the crossover region, the so-called pseudogap phenomenon is expected\cite{Tsuchiya}, where strong pairing fluctuations induce a gap-like structure in single-particle excitation spectra, in spite of the normal state. 
\par
In considering strong-coupling effects on a $p$-wave interacting normal Fermi gas, one naive idea is to examine the pseudogap phenomenon appearing in single-particle excitations, by observing the single-particle density of states $\rho(\omega)$, as well as the single-particle spectral weight $A({\bm p},\omega)$. Although this approach seems promising at a glance, Ref.\cite{Inotani} pointed out that the pseudogap structure appearing in $\rho(\omega)$ and $A({\bm p},\omega)$ is not so remarkable as the $s$-wave case. This is because, as one approaches the intermediate coupling regime from the weak-coupling side, the Fermi chemical potential $\mu$ remarkably decreases from the Fermi energy $\varepsilon_{\rm F}$\cite{Gurarie,Ohashi,Ho,Botelho,Iskin,Iskin2,Cheng}. Then, the $p$-wave interaction strength around the momentum $p=\sqrt{2m\mu}$ (where $m$ is an atomic mass), where the pseudogap is expected in the spectral weight $A({\bm p},\omega)$ becomes weak as, symbolically, $p^2U_p\sim 2m\mu U_p$ (where $U_p$ is the $p$-wave coupling constant, and the factor $p^2$ comes from the momentum dependence of a $p$-wave pairing interaction). As a result, the pseudogap regime where the pseudogap is seen in single-particle excitations vanishes in the intermediate coupling regime where $\mu\sim 0$\cite{Inotani}, in spite of the fact that the $p$-wave scattering volume almost diverges there.
\par
Keeping this in mind, this paper takes another strategy. That is, we approach normal-state properties of a $p$-wave interacting Fermi gas, on the viewpoint of the specific heat $C_V$ at constant volume. Including fluctuations in the $p$-wave Cooper channel within the framework of the strong-coupling theory developed by Nozi\`eres and Schmitt-Rink (NSR)\cite{Inotani,NSR}, we calculate the temperature dependence of this thermodynamic quantity in the whole coupling regime above $T_{\rm c}$. Since the temperature dependence of $C_V$ is very different between a free Fermi gas and a Bose gas, this thermodynamic quantity is convenient to identify the ``Fermi atomic gas regime" and "molecular Bose gas regime" in the phase diagram of a $p$-wave interacting Fermi gas. In addition, $C_V$ is also influenced by pairing fluctuations because it is deeply related to the entropy $S$, so that it can also be used to identify the region where strong $p$-wave pairing fluctuations exist. We briefly note that Ref. \cite{Pieter} has recently used these advantages of $C_V$ to investigate normal-state properties of an $s$-wave interacting Fermi gas\cite{Pieter}, to successfully obtain the phase diagram of this system, consisting of (1) the normal Fermi gas regime, (2) the molecular Bose gas regime, (3) the pseudogap regime, in addition to (4) the superfluid phase. We also note that the observation of the specific heat has recently become possible in cold Fermi gas physics\cite{Ku}.
\par
In a sense, the present work is an extension of our previous work\cite{Pieter} for an $s$-wave interacting Fermi gas to the $p$-wave case. However, as pointed out in Ref.\cite{Inotani3}, one should recall that a $p$-wave pairing interaction causes an anomalous behavior of $C_V$ in relatively high temperature region of the weak-coupling regime. (Note that this phenomenon is absent in the $s$-wave case.) Thus, in addition to effects of pairing fluctuations, it is also an interesting issue how this anomaly influences our attempt that we construct the phase diagram of a $p$-wave interacting Fermi gas on the viewpoint of $C_V$.
\par
This paper is organized as follows. In Sec. II, we explain our formulation based on the strong-coupling NSR theory. We also explain how to calculate the specific heat $C_V$ at constant volume. In Sec. III, we discuss how $p$-wave pairing fluctuations affect the specific heat $C_V$ at $T_{\rm c}$, from the weak-coupling regime to the strong-coupling regime. In Sec. IV, we examine the temperature dependence of $C_V$ above $T_{\rm c}$. Using this result, we draw the phase diagram of a $p$-wave interacting Fermi gas in terms of the temperature and the $p$-wave interacting strength. Throughout this paper, we set $\hbar=k_{\rm B}=1$, and the system volume $V$ is taken to be unity, for simplicity.
\par
%%%%%%%%%%%%%%%%%%%%%%%%%%%%%%%%%%%%%%%%%%%%%%%%%%%%%%%%%%%%%%%%%%%%%%%%%%%%%%%
\section{Formulation}
\par
We consider a one-component Fermi gas with a $p$-wave pairing interaction, described by the Hamiltonian,
\begin{equation}
H=\sum_{\bm p} \xi_{\bm p}c_{\bm p}^{\dagger}c_{\bm p}
-\frac{1}{2}\sum_{{\bm p},{\bm p}',{\bm q}} 
V_p({\bm p},{\bm p}')
c_{{\bm p}+{\bm q}/2}^\dagger c_{-{\bm p}+{\bm q}/2}^\dagger
c_{-{\bm p}'+{\bm q}/2}c_{{\bm p}'+{\bm q}/2}.
\label{eq.1}
\end{equation}
Here, $c^\dagger_{\bm p}$ is a creation operator of a Fermi atom with the kinetic energy $\xi_{\bm p}=p^2/(2m)-\mu$, measured from the Fermi chemical potential $\mu$ (where $m$ is an atomic mass). $V_p({\bm p},{\bm p}')$ is a $p$-wave pairing interaction, having the form\cite{Ohashi,Ho},
\begin{equation}
V_p({\bm p},{\bm p}')=-\sum_{i=x,y,z} \gamma^i_{\bm p} U_p \gamma^i_{{\bm p}'},
\label{eq.2}
\end{equation}
where the coupling constant $-U_p~(<0)$ is assumed to be tunable by a Feshbach resonance. The $p$-wave symmetry is characterized by the basis function, $\gamma^i_{\bm p}=p_i F_{\rm c}({\bm p})~(i=x,y,z)$, where $F_{\rm c}({\bm p})=1/[1+(p/p_{\rm c})^6]$ is a cutoff function\cite{Inotani3,note2}, with $p_{\rm c}$ being a cutoff momentum. The $p$-wave interaction in Eq. (\ref{eq.2}) is ``isotropic" in the sense that all the three $p_i$-wave components ($i=x,y,z$) have the same coupling strength $U_p$. In this regard, we note that an anisotropic $p$-wave interaction, $V_p({\bm p},{\bm p}')=-\sum_{i=x,y,z} \gamma^i_{\bm p} U^i_p \gamma^i_{{\bm p}'}$ ($U_p^x>U_p^y=U_p^z$), has been discovered in a $^{40}$K Fermi gas\cite{note}. However, for simplicity, we ignore this uni-axial anisotropy, effects of which will be separately discussed in our future paper. As usual, we measure the $p$-wave interaction strength in terms of the inverse scattering volume $v^{-1}$, which is related to the bare interaction $U_p$ as\cite{Ho},
\begin{equation}
{4\pi v \over m}=-
{U_p \over 3}
{1 \over \displaystyle 1-{U_p \over 3}\sum_{\bm p}{p^2 \over 2\varepsilon_{\bm p}}F_{\rm c}^2({\bm p})},
\label{eq.3}
\end{equation}
where the cutoff momentum $p_{\rm c}$ in $F_{\rm c}({\bm p})$ is related to the inverse effective range $k_0$ as
\begin{equation}
k_0=-{4\pi \over m^2}
\sum_{\bm p}{{\bm p}^2 \over 2\varepsilon_{\bm p}^2}F_{\rm c}^2({\bm p}).
\label{eq.4}
\end{equation}
Following the experiment on a $^{40}$K Fermi gas\cite{Ticknor}, we set $k_0=-30k_{\rm F}$ (where $k_{\rm F}$ is the Fermi momentum). When we use the scattering volume $v$, the weak-coupling side and the strong-coupling side are conveniently characterized as $(k_{\rm F}^3v)^{-1}\lesssim 0$ and $(k_{\rm F}^3 v)^{-1}\gesim 0$, respectively.
\par
%%%%%%%%%%%%%%%%%%%%%%%%%%%%%%%%%%%%%%%%%%%%%%%%%%%%%%%%%%%%%%%%%%%%%%%%%%%%%%%
\begin{figure}
\centerline{\includegraphics[width=8cm]{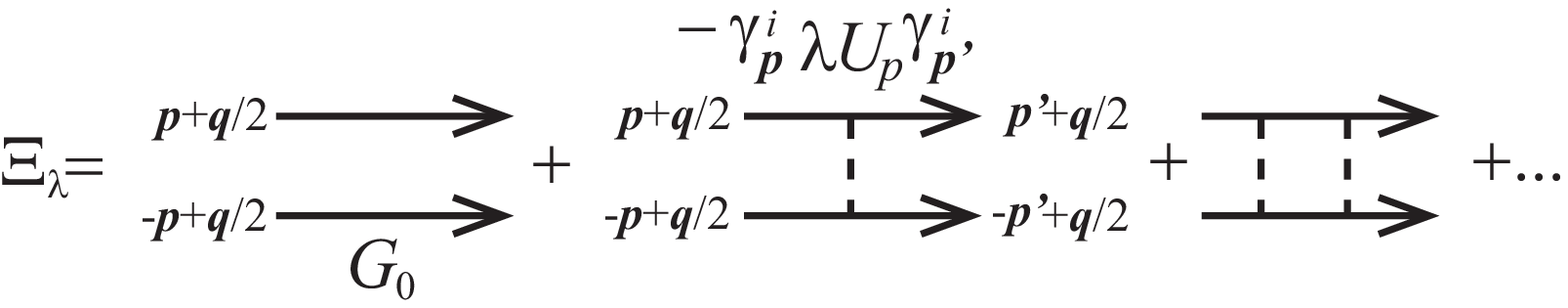}}
\caption{ NSR particle-particle scattering matrix $\Xi_\lambda({\bm q},i\nu_n)$ in Eq. (\ref{eq.8}). The solid line and dashed line represent the bare single-particle thermal Green's function $G_0({\bm p}, i\omega_n)=[i\omega-\xi_{\bm p}]^{-1}$ and the $p$-wave interaction $V_p({\bm p},{\bm p}')$, respectively. $\omega_n$ is the fermion Matsubara frequency.}
\label{fig1}
\end{figure}
%%%%%%%%%%%%%%%%%%%%%%%%%%%%%%%%%%%%%%%%%%%%%%%%%%%%%%%%%%%%%%%%%%%%%%%%%%%%%%%
\par
The specific heat $C_V$ at constant volume is conveniently calculated from
\begin{equation}
C_V=
\left(
{\partial E \over \partial T}
\right)_{V,N},
\label{eq.5cv}
\end{equation}
where the internal energy $E$ is related to the thermodynamic potential $\Omega$ as (Legendre transformation),
\begin{equation}
E=
\Omega
-T\left( \frac{\partial \Omega}{\partial T} \right)_{\mu}
-\mu\left( \frac{\partial \Omega }{\partial \mu} \right)_{T}.
\label{eq.5}
\end{equation}
\par
In this paper, we take into account strong-coupling corrections to $\Omega$ within the framework of the NSR strong-coupling theory\cite{NSR}. Replacing the $p$-wave coupling constant $U_p$ by $\lambda U_p$, we have\cite{AGD}
\begin{eqnarray}
{\partial \Omega \over \partial \lambda}
&=&
{U_p \over 2}
\sum_{{\bm p},{\bm p}',{\bm q}}\sum_{i=x,y,z}
\gamma^i_{\bm p}\gamma^i_{{\bm p}'}
\langle
c_{{\bm p}+{\bm q}/2}^\dagger c_{-{\bm p}+{\bm q}/2}^\dagger
c_{-{\bm p}'+{\bm q}/2}c_{{\bm p}'+{\bm q}/2}
\rangle
\nonumber
\\
&=&
U_pT\sum_{{\bm q},\nu_n}\Xi_\lambda({\bm q},i\nu_n).
\label{eq.6}
\end{eqnarray}
Here, $\Xi_\lambda({\bm q},i\nu_n)$ is the particle-particle scattering matrix for the $p$-wave interaction $\lambda V_p({\bm p},{\bm p}')$ (where $\nu_n$ is the boson Matsubara frequency), having the form
\begin{eqnarray}
\Xi_\lambda({\bm q},i\nu_n)
&=&
{1 \over 2}
\sum_{{\bm p},{\bm p'}}\sum_{i=x,y,z}
\gamma^i_{\bm p}\gamma^i_{{\bm p}'}
\int_0^{1/T}d\tau
e^{i\nu_n\tau}
\nonumber
\\
&{}&
\hskip-20mm
\times
\langle T_\tau
\{
c_{{\bm p}+{\bm q}/2}^\dagger(\tau) c_{-{\bm p}+{\bm q}/2}^\dagger(\tau)
c_{-{\bm p}'+{\bm q}/2}(0)c_{{\bm p}'+{\bm q}/2}(0)
\}
\rangle,
\label{eq.7}
\end{eqnarray}
where $A(\tau)=e^{\tau H}Ae^{-\tau H}$. The NSR thermodynamic potential $\Omega$ is obtained by evaluating $\Xi_\lambda({\bm q},i\nu_n)$ in Eq. (\ref{eq.7}) within the $T$-matrix (ladder) approximation with respect to the $p$-wave interaction $\lambda V_p({\bm p},{\bm p}')$, which is diagrammatically given as Fig. \ref{fig1}. Summing up these diagrams, one has,
\begin{eqnarray}
\Xi_\lambda({\bm q},i\nu_n)
&=&{\rm {Tr}} \left[
{2\hat{\Pi}({\bm q},i\nu_n) \over 1-\lambda U_p\hat{\Pi}({\bm q},i\nu_n)}
\right],
\label{eq.8}
\end{eqnarray}
where
\begin{eqnarray}
\Pi_{ij}({\bm q},i\nu_n)
=
-\sum_{\bm {p}} 
\gamma^i_{\bm p} 
\gamma^j_{\bm p}
\frac{1-f(\xi_{\bm{p}+{\bm q}/2})-f(\xi_{-\bm{p}+{\bm q}/2})}
{i\nu_n - \xi_{\bm{p}+{\bm q}/2} - \xi_{-\bm{p}+{\bm q}/2}}
\label{eq.9}
\end{eqnarray}
is the $p$-wave pair correlation function, with $f(x)$ being the Fermi distribution function. Substituting Eq. (\ref{eq.8}) into Eq. (\ref{eq.6}), and then carrying out the integration over $\lambda$ from $\lambda=0$ to $\lambda=1$, we reach the NSR thermodynamic potential\cite{Inotani,Inotani2,Ohashi,Ho,Botelho,Iskin,Iskin2},
\begin{equation}
\Omega=\Omega_0+T\sum_{{\bm q},i\nu_n}
{\rm {Tr}}
\ln \left[1-U_p\hat{\Pi}({\bm q},i\nu_n)\right],
\label{eq.10}
\end{equation}
where
\begin{equation}
\Omega_0=T \sum_{\bm p} \ln \left[1+e^{-\xi_{\bm p}/T} \right]
\label{eq.11_0}
\end{equation}
is the thermodynamic potential in a free Fermi gas. 
\par
Once the NSR thermodynamic potential $\Omega$ is obtained, the internal energy $E$ in Eq. (\ref{eq.5}) can immediately be evaluated as
\begin{eqnarray}
E
&=&
\sum_{{\bm p}}\varepsilon_{\bm p} f(\xi_{\bm p})
\nonumber
\\
&-&
T\sum_{ {\bm q},i\nu_n}
{\rm {Tr}}\left[
\hat{\Gamma}({\bm q},i\nu_n)
\left(
T\frac{ \partial \hat{\Pi}({\bm q},i\nu_n)}{\partial T}
+\mu\frac{\partial \hat{\Pi}({\bm q},i\nu_n)}{\partial \mu}
\right)
\right].
\nonumber
\\
\label{eq.11}
\end{eqnarray}
where $\hat{\Gamma}({\bm q},i\nu_n)=U_p[1-U_p\hat{\Pi}({\bm q},i\nu_n)]^{-1}$.
\par
Before ending this section, we explain detailed computations. We numerically evaluate the derivative in Eq. (\ref{eq.5cv}), by calculating the NSR internal energy $E$ in Eq. (\ref{eq.11}) at slightly different two temperatures. In this procedure, we also need to determine the Fermi chemical potential $\mu$, which is, as usual, achieved by considering the equation for the number $N$ of Fermi atoms,\cite{NSR},
\begin{eqnarray}
N
&=&
-\left(
{\partial \Omega \over \partial\mu}
\right)_T
\nonumber
\\
&=&
\sum_{\bm p}f(\xi_{\bm p})
+T \sum_{{\bm q},\nu_n}
{\rm {Tr}}
\left[ 
\hat{\Gamma}({\bm q},i\nu_n)
\left( \frac{\partial \hat{\Pi}({\bm q},i\nu_n)}{\partial \mu} \right)
\right].
\nonumber
\\
\label{eq.12}
\end{eqnarray}
The $p$-wave superfluid phase transition temperature $T_{\rm c}$ is determined by solving the number equation (\ref{eq.12}), imposing the Thouless criterion, $1=U_p\Pi_{ii}({\bm q}=0,i\nu_n=0)$, which gives,
\begin{eqnarray}
1={U_p \over 3}\sum_{\bm p}F_{\rm c}^2({\bm p}) {{\bm p}^2 \over 2\xi_{\bm p}}\tanh \left( \frac{\xi_p}{2T} \right).
\label{eq.13}
\end{eqnarray}
\par
%%%%%%%%%%%%%%%%%%%%%%%%%%%%%%%%%%%%%%%%%%%%%%%%%%%%%%%%%%%%%%%%%%%%%%%%%%%%%%%
\begin{figure}
\centerline{\includegraphics[width=8cm]{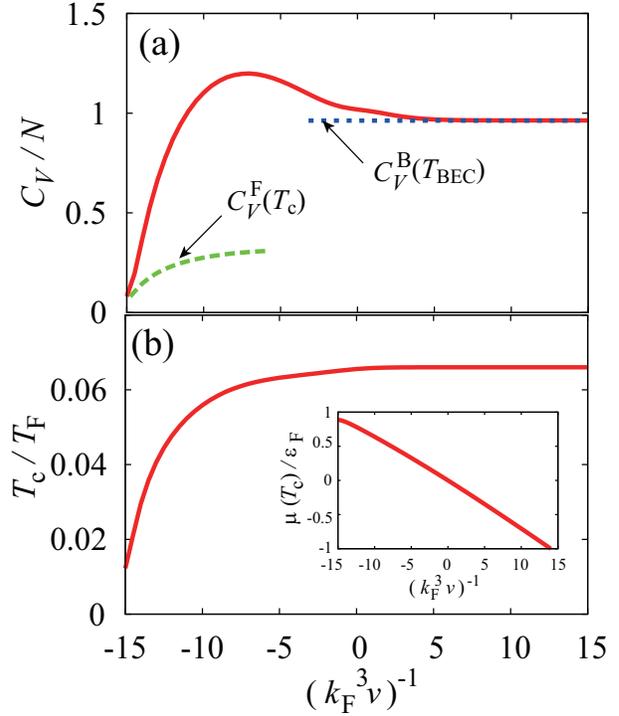}}
\caption{(Color online) (a) Calculated specific heat $C_V$ at constant volume in a $p$-wave interacting Fermi gas at $T_{\rm c}$. $C_V^{\rm F}(T_{\rm c}$ is the specific heat of a free Fermi gas at $T_{\rm c}$. $C_V^{\rm B}(T_{\rm BEC})=0.937N$ is the specif heat at the BEC phase transition temperature $T_{\rm BEC}$ in an ideal Bose gas mixture, consisting of three kinds of $N/6$ $p_i$-wave molecules ($i=x,y,z$). (b) Superfluid phase transition temperature $T_{\rm c}$. The inset shows the Fermi chemical potential $\mu(T=T_{\rm c})$. $T_{\rm F}$ and $\varepsilon_{\rm F}$ are the Fermi temperature and Fermi energy, respectively.}
\label{fig2}
\end{figure}
%%%%%%%%%%%%%%%%%%%%%%%%%%%%%%%%%%%%%%%%%%%%%%%%%%%%%%%%%%%%%%%%%%%%%%%%%%%%%%%
\par
\section{Specific heat $C_V$ at the $p$-wave superfluid transition temperature}
\par
Figure \ref{fig2}(a) shows the calculated specific heat $C_V$ at constant volume in a $p$-wave interacting Fermi gas at $T_{\rm c}$. In this figure, we find that $C_V(T_{\rm c})$ is remarkably enhanced around $(k_{\rm F}v)^{-1}\simeq -7.5$. In the weaker-coupling regime, $C_V(T_{\rm c})$ is reduced to the specific heat $C_V^{\rm F}(T_{\rm c})$ in a free Fermi gas. On the hand, $C_V(T_{\rm c})$ approaches a constant value, when $(k_{\rm F}v)^{-1}\gesim 5$. In the latter strong-coupling regime, the Fermi chemical potential $\mu$ is negative, and the superfluid phase transition temperature $T_{\rm c}$ is insensitive to the interacting strength, as shown in Fig. \ref{fig2}(b), indicating that most Fermi atoms form tightly bound molecules with a large binding energy ($E_{\rm bind} \sim 2\mu(T_{\rm c})$). Indeed, noting that each $p_i$-wave Cooper channel ($i=x,y,z$) has $N/6$ molecules in the strong-coupling limit, we can evaluate $T_{\rm c}$ in this limiting case by simply calculating the BEC phase transition temperature $T_{\rm BEC}$ in this ideal Bose gas mixture, given by\cite{Ohashi},
\begin{equation}
T_{\rm BEC}={T_{\rm F} \over 3[6\sqrt{\pi}\zeta(3/2)]^{2/3}}=0.066T_{\rm F}. 
\label{eq.15}
\end{equation}
This value agrees well with $T_{\rm c}$ in Fig. \ref{fig2}(b) when $(k_{\rm F}v)^{-1}\gesim 5$. In addition, the specific heat $C_V^{\rm B}(T_{\rm BEC})$ in this Bose mixture, 
\begin{eqnarray}
C_V^{\rm B}(T_{\rm BEC}) = 
{45 \over 4}{\zeta(5/2) \over \zeta(3/2)} N_{\rm B}=0.963N,
\label{eq.14}
\end{eqnarray}
also explains the behavior of $C_V(T_{\rm c})$ in the strong-coupling regime as shown in Fig. \ref{fig2}(a) (where $\zeta(3/2)=2.612$ and $\zeta(5/2)=1.341$ are zeta functions).
\par
Since we are only including fluctuations in the $p$-wave Cooper channel, the enhancement of $C_V(T_{\rm c})$ seen in Fig. \ref{fig2}(a) around $(k_{\rm F}v)^{-1}\simeq -7.5$ is considered as a many-body fluctuation phenomenon. (Note that $\mu(T_{\rm c})>0$ in this regime (see the inset in Fig. \ref{fig2}(b)), so that there is no two-body bound molecule in this weak-coupling region.) To understand this phenomenon in more detail, it is convenient to consider the NSR single-particle thermal Green's function $G({\bm p},i\omega_n)$, given by\cite{Fukushima}
\begin{eqnarray}
G({\bm p},i\omega_n)
&=&
G_0({\bm p},i\omega_n)+
G_0({\bm p},i\omega_n)\Sigma({\bm p},i\omega_n)G_0({\bm p},i\omega_n)
\nonumber
\\
&\simeq&
{1 \over i\omega_n-\xi_{\bm p}-\Sigma({\bm p},i\omega_n)}.
\label{eq.16}
\end{eqnarray}
Here, $G_0({\bm p},i\omega_n)=[i\omega_n-\xi_{\bm p}]^{-1}$ is the bare Green's function, where $\omega_n$ is the fermion Matsubara frequency. We briefly note that the expression in the first line in Eq. (\ref{eq.16}) is obtained so as to reproduce the NSR number equation (\ref{eq.12})\cite{Fukushima}. Noting that the NSR particle-particle scattering matrix $\Gamma_{\lambda=1}({\bm q},i\nu_n)$ in Eq. (\ref{eq.7}) is enhanced around ${\bm q}=\nu_n=0$ near $T_{\rm c}$, one may approximate the self-energy correction $\Sigma({\bm q},i\nu_n)$ in Eq. (\ref{eq.16}) to
\begin{eqnarray}
\Sigma({\bm p},i\omega_n)
&=&
T\sum_{{\bm q},\nu_n,i,j}\gamma_{{\bm p}-{\bm q}/2}^i
\Gamma_{ij}({\bm q},i\nu_n)\gamma_{{\bm p}-{\bm q}/2}^j
\nonumber
\\
&\times& G_0({\bm q}-{\bm p},i\nu_n-i\omega_n)
\nonumber
\\
&\simeq&
\Delta_{\rm pg}^2({\bm p})G_0(-{\bm p},-i\omega_n),
\label{eq.17}
\end{eqnarray}
In the last expression, 
\begin{equation}
\Delta_{\rm pg}^2({\bm p})=T{\bm p}^2\sum_{{\bm q},\nu_n} {\rm {Tr}}\left[ \hat{\Gamma}({\bm q},i\nu_n)\right]~~~(>0)
\label{eq.17b}
\end{equation}
physically describes a particle-hole coupling caused by strong $p$-wave pairing fluctuations, which is also referred to as the pseudogap parameter in the literature\cite{Levin}. Substituting the last expression in Eq. (\ref{eq.17}) into the second line in Eq. (\ref{eq.16}), the pseudogap parameter $\Delta_{\rm pg}$ is found to combine the particle Green's function $G_0({\bm p},i\omega)=[i\omega-\xi_{\bm p}]^{-1}$ with the hole Green's function $G^{\rm hole}_0({\bm p},i\omega)=[i\omega+\xi_{\bm p}]^{-1}$ as,
\begin{equation}
G({\bm p},i\omega_n)=
{1 \over 
\displaystyle 
G_0^{-1}({\bm p},i\omega_n)-
\Delta_{\rm pg}^2 G_0^{\rm hole}({\bm p},i\omega_n)}.
\label{eq.17c}
\end{equation}
This single-particle Green's function has the same form as the diagonal component of the BCS Green's function in the superfluid state as,
\begin{equation}
G({\bm p},i\omega_n)=-
{i\omega_n+\xi_{\bm p} \over \omega_n^2+\xi_{\bm p}^2+\Delta_{\rm pg}^2}.
\label{eq.18}
\end{equation}
This means that a normal Fermi gas has superfluid-like properties near $T_{\rm c}$, when $p$-wave pairing fluctuations described by the particle-particle scattering matrix $\Gamma_{\lambda=1}({\bm q},i\nu_n)$ are enhanced in the low-energy and low-momentum region (${\bm q}=\nu_n=0$). This causes the suppression of the entropy $S$ near $T_{\rm c}$ as in the superfluid phase below $T_{\rm c}$, leading naturally to the enhancement of the specific heat at constant volume,
\begin{equation}
C_V=T\left(\partial S \over \partial T\right)_{V,N},
\label{eq.18b}
\end{equation}
compared to the case of a free Fermi gas. 
\par
\begin{figure}
\centerline{\includegraphics[width=8cm]{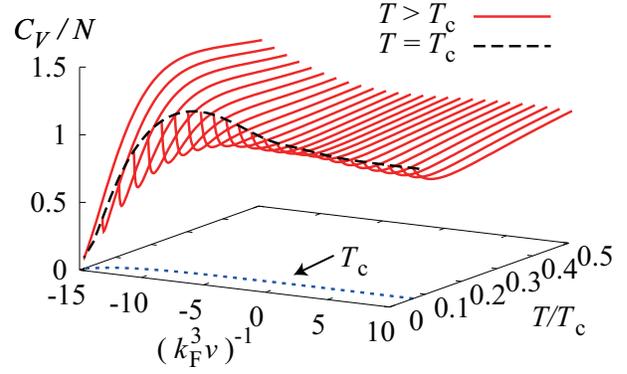}}
\caption{(Color online) Calculated specific heat $C_V$ at constant volume above $T_{\rm c}$, at various $p$-wave interaction strengths. The dashed line shows $C_V(T=T_{\rm c})$.}
\label{fig3}
\end{figure}

According to the above-mentioned ``pairing-fluctuation" scenario, the remarkable enhancement of $C_V(T_{\rm c})$ seen in Fig. \ref{fig2}(a) around $(k_{\rm F}v)^{-1}\simeq -7.5$ should only occur near $T_{\rm c}$ where low-energy and low-momentum fluctuations in the $p$-wave Cooper channel are strong. This can be confirmed by examining $C_V$ above $T_{\rm c}$, as shown in Fig. \ref{fig3}. 
\par
%%%%%%%%%%%%%%%%%%%%%%%%%%%%%%%%%%%%%%%%%%%%%%%%%%%%%%%%%%%%%%%%%%%%%%%%%%%%%%%
%%%%%%%%%%%%%%%%%%%%%%%%%%%%%%%%%%%%%%%%%%%%%%%%%%%%%%%%%%%%%%%%%%%%%%%%%%%%%%%
\par
\section{Phase diagram of a $p$-wave interacting Fermi gas on the viewpoint of specific heat $C_V$}
\par
\begin{figure}
\centerline{\includegraphics[width=8cm]{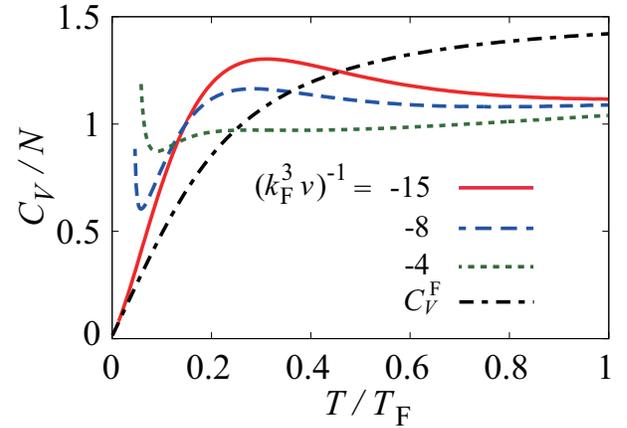}}
\caption{(Color online) Calculated specific heat $C_V$ as a function of temperature, in the weak coupling side ($(k_{\rm F}^3 v)^{-1} <0$). $C_V^{\rm F}$ is the specific heat in an free Fermi gas.}
\label{fig4}
\end{figure}
\begin{figure}
\centerline{\includegraphics[width=8cm]{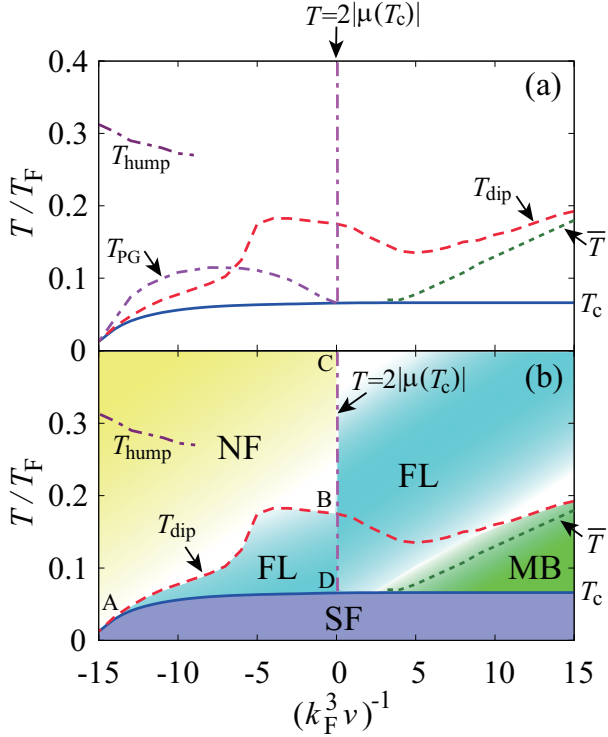}}
\caption{(Color online) (a) Characteristic temperature $T_{\rm dip}$ determined  as the temperature at the dip structure in the temperature dependence of the specific heat $C_V$. $T_{\rm PG}$ is the pseudogap temperature which is determined as the temperature below which a dip structure appears around $\omega=0$ in the single-particle density of states $\rho(\omega)$\cite{Inotani}. $T_{\rm hump}$ is the temperature at the top of the hump structure in $C_V$. ${\bar T}$ is the temperature satisfying $C_V(T)/C_V^{\rm B}(T)=1.01$, where $C_V^{\rm B}(T)$ is the specific heat in an ideal Bose gas with three kinds of $N/6$ $p$-wave molecules. We also draw the characteristic temperature satisfying $T=2|\mu(T)|$ in the strong-coupling side where $\mu<0$, below which two-body bound molecules starts to appear, overwhelming thermal dissociation. (b) Phase diagram of a one-component Fermi gas with a $p$-wave pairing interaction. SF: $p$-wave superfluid phase. NF: normal Fermi gas regime (although $C_V$ is still affected by $p$-wave interaction, exhibiting a hump structure in the temperature dependence). FL: the region where pairing fluctuations are important. MB: molecular Bose gas regime. 
}
\label{fig5}
\end{figure}

Extracting $C_V(T\ge T_{\rm c})$ from the weak-coupling side ($(k_{\rm F}^3 v)^{-1}<0$) in Fig. \ref{fig3}, we obtain Fig. \ref{fig4}, where $p$-wave pairing fluctuations are found to give a dip structure in the temperature dependence of this thermodynamic quantity. Thus, although there is no phase transition at the dip, the ``dip temperature" $T_{\rm dip}$ is expected to work as a characteristic temperature in this regime, below which strong $p$-wave pairing fluctuations affect system properties, such as the specific heat $C_V$. Indeed, as shown in Fig. \ref{fig5}(a), $T_{\rm dip}$ in the weak-coupling side is comparable to the previous pseudogap temperature $T^*$\cite{Inotani} (which is determined as the temperature below which the single-particle density of states $\rho(\omega)$ has a dip structure around $\omega=0$). Although they do not have to completely coincide with each other because the both are crossover temperatures without being accompanied by any phase transition, this result makes us expect that $T_{\rm dip}$ can be used to roughly estimate the pseudogap temperature $T^*$ from the observation of $C_V$, within the accuracy shown in Fig. \ref{fig5}(a).
\par
%%%%%%%%%%%%%%%%%%%%%%%%%%%%%%%%%%%%%%%%%%%%%%%%%%%%%%%%%%%%%%%%%%%%%%%%%%%%%%
%%%%%%%%%%%%%%%%%%%%%%%%%%%%%%%%%%%%%%%%%%%%%%%%%%%%%%%%%%%%%%%%%%%%%%%%%%%%%%
\par
%%%%%%%%%%%%%%%%%%%%%%%%%%%%%%%%%%%%%%%%%%%%%%%%%%%%%%%%%%%%%%%%%%%%%%%%%%%%%%
%%%%%%%%%%%%%%%%%%%%%%%%%%%%%%%%%%%%%%%%%%%%%%%%%%%%%%%%%%%%%%%%%%%%%%%%%%%%%%
\par
However, quantitatively, one sees in Fig. \ref{fig5}(a) that $T_{\rm dip}<T^*$ when $(k_{\rm F}^3v)^{-1}\lesssim -7$. In this regard, we note that $C_V(T>T_{\rm dip})$ in this regime exhibits a hump structure (see around $T/T_{\rm F}=0.3$ in Fig. \ref{fig4})\cite{Inotani3}, originating from anomalous particle-particle scatterings into virtual $p$-wave molecular states. Because of this anomaly, $C_V(T>T_{\rm dip})$ does not coincide with the specific heat $C_V^{\rm F}(T)$ in a free Fermi gas, as shown in Fig. \ref{fig4}. Even when $(k_{\rm F}^3 v)^{-1}=-15$ (where the enhancement by $p$-wave pairing fluctuations is not seen at all), Fig. \ref{fig4} shows that $C_V(T)$ still deviates from $C_V^{\rm F}(T)$, except at very low temperatures. This is quite different from the $s$-wave case, where $C_V(T>T_{\rm dip})$ is well described by a free Fermi gas\cite{Pieter}. Because of this hump structure in the $p$-wave case, the dip position in the region $(k_{\rm F}v)^{-1}\lesssim -7$ is considered to be lowered to some extent, compared to the case when such a hump structure is absent and the region above $T_{\rm dip}$ is simply described by a free Fermi gas, which may be a reason for $T_{\rm dip}<T^*$ in this region. 
\par
Indeed, around $(k_{\rm F}^3v)^{-1}=-5$ (where the hump no longer exists, see $T_{\rm hump}$ in Fig. \ref{fig5}(a)), Fig. \ref{fig5}(a) shows that $T_{\rm dip}$ remarkably increases with increasing the interaction strength, to exceeds the pseudogap temperature $T^*$. As mentioned previously, the pseudogap phenomenon in the density of states is suppressed in the intermediate coupling regime ($(k_{\rm F}^3v)^{-1}\sim 0$), because of a combined effects of small Fermi chemical potential $\mu$ with the momentum dependence of the $p$-wave interaction $V_p({\bm p},{\bm p}')\propto{\bm p}\cdot{\bm p}'$. As a result, although the scattering volume is almost diverges there, the pseudogap temperature $T^*$ vanishes at $(k_{\rm F}^3 v)^{-1}\simeq 0$ (where $\mu(T_{\rm c})\sim 0$). Since such an effect is absent in the specific heat, one obtains $T_{\rm dip}>T^*$ there. Thus, in this regime, $T_{\rm dip}$ would be more useful than $T^*$, in examining the region where $p$-wave pairing fluctuations are strong.
\par
%%%%%%%%%%%%%%%%%%%%%%%%%%%%%%%%%%%%%%%%%%%%%%%%%%%%%%%%%%%%%%%%%%%%%%%%%%%%%%
\begin{figure}
\centerline{\includegraphics[width=8cm]{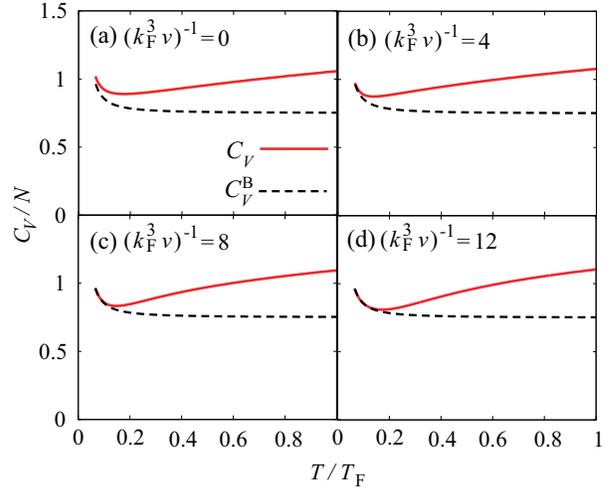}}
\caption{(Color online) Same as Fig. \ref{fig4} for the strong-coupling side ($(k_{\rm F}^3 v)^{-1} \ge 0$). $C_V^{\rm B}$ is the specific heat in an ideal Bose gas with three kinds of $N/6$ molecules.}
\label{fig6}
\end{figure}
%%%%%%%%%%%%%%%%%%%%%%%%%%%%%%%%%%%%%%%%%%%%%%%%%%%%%%%%%%%%%%%%%%%%%%%%%%%%%
\par
Figure \ref{fig5}(a) shows that $T_{\rm dip}$ is also obtained in the strong-coupling side ($(k_{\rm F}v)^{-1}>0$). However, we note that the physical meaning of $T_{\rm dip}$ in this regime is quite different from the weak-coupling case. To see this, Fig. \ref{fig6} compares $C_V(T)$ with the specific heat $C_V^{\rm B}$ in an ideal Bose gas mixture, consisting of three kinds of $N/6$ $p$-wave molecules. Near $T_{\rm c}$, one sees that $C_V(T)$ agrees well with $C_V^{\rm B}(T)$, indicating that the system is very close to this molecular Bose gas. The fact that $C_V(T)$ gradually deviates from $C_V^{\rm B}(T)$ with increasing the temperature is simply because of the onset of the thermal dissociation of these molecules into Fermi atoms. 
\par
To characterize this thermal dissociation in a quantitative manner, when we conveniently plot the temperature ${\bar T}$ at which $C_V(T)/C_V^{\rm B}(T)=1.01$ is satisfied\cite{notezz}, it is found to become close to the dip temperature $T_{\rm dip}$ with increasing the interaction strength, especially when $(k_{\rm F}^3v)^{-1}\gesim 10$. Thus, $T^{\rm dip}$ in this regime may be interpreted as the characteristic temperature which distinguishes between the region where the system may be viewed as an ideal Bose gas of tightly bound $p$-wave molecules ($T<T_{\rm dip}~(\sim {\bar T})$), and the region where some of them are thermally dissociated into Fermi atoms ($T>T_{\rm dip}~(\sim{\bar T})$). 
\par
According to the above classification, (thermal) pairing fluctuations are important in the latter region ($T>T_{\rm dip}$). In this regard, we note that this is quite different from the weak-coupling case, where pairing fluctuations become crucial {\it below} $T_{\rm dip}$. Because of this, although $T_{\rm dip}$ has the clear physical meaning in the weak-coupling side ($(k_{\rm F}^3v)^{-1}<0$), as well as in the strong-coupling regime ($(k_{\rm F}^3v)^{-1}\gesim 10)$, it is difficult to give a clear physical picture to $T_{\rm dip}$ in the region $0\lesssim (k_{\rm F}^3v)^{-1}\lesssim 10$, which remains as our future problem.
\par
Since the physical meaning of $T_{\rm dip}$ in the strong-coupling regime is different from that in the weak-coupling side, it is interesting to explore another characteristic temperature in the former regime that has a similar physical meaning to $T_{\rm dip}$ in the weak-coupling side. In this regard, we point out that the Fermi chemical potential $\mu$ may be useful. To explain this, we recall that $\mu$ becomes negative when $(k_{\rm F}^3v)^{-1}\gesim 0$ (see the inset in Fig. \ref{fig2}(b))\cite{Ohashi,Ho}. Then, since $2\mu$ has the meaning of the energy which needs to add two Fermi atoms to the system, a negative $\mu$ indicates that a bound molecules with the binding energy $2\mu$ are formed, when two fermions are introduced to the system. Indeed, in the extreme BEC limit, it has been shown that $2\mu$ is reduced to the binding energy of a two-body bound state\cite{Ho}, given by
\begin{equation}
E_{\rm bind}=-{2 \over m|k_0|v}~~~(<0).
\label{eq.19}
\end{equation}
Thus, when we consider the temperature which satisfies
\begin{equation}
T=2|\mu(T_{\rm c})|~~~(\mu<0),
\label{eq.20}
\end{equation}
it physically has the meaning that two-body bound molecules start to form below  around this temperature, overwhelming thermal dissociation. As shown in Fig. \ref{fig5}(a), Eq. (\ref{eq.20}) gives an almost vertical line at $(k_{\rm F}^3v)^{-1}\simeq 0$ (although it is actually an increase function of $(k_{\rm F}^3v)^{-1}$). Then, the region between this line and ${\bar T}$ may be regarded as the regime where fluctuating molecular bosons are dominant, which corresponds to the pairing-fluctuation regime below $T_{\rm dip}$ in the weak-coupling side. 
\par
Using the above discussion, we obtain the phase diagram of a one-component Fermi gas with a $p$-wave interaction shown in Fig. \ref{fig5}(b). In the normal state above $T_{\rm c}$, the region ``FL" between the line ABC and ${\bar T}$ are characterized by strong pairing fluctuations. In this regime, the left side of the line BD is dominated by fluctuations of preformed $p$-wave Cooper pairs. The right side of the line BD is dominated by two-body bound molecules that are partially dissociated into Fermi atoms by thermal effects. In the region ``MB", thermal dissociation of these molecules are almost absent, so that the system is well described by an ideal Bose gas with three-kinds of $N/6$ $p$-wave molecules. Pairing fluctuations are weak in the ``normal-Fermi gas region (NF)", although the $p$-wave interaction still affects the specific heat $C_V$, giving a hump structure in the temperature dependence. 
\par
We emphasize that $T_{\rm c}$ is only the phase transition temperature in Fig. \ref{fig5}(b). The others are all crossover temperatures without being accompanied by any phase transition. However, Fig. \ref{fig5}(b) would be still useful in considering how the normal-state properties of a $p$-wave interacting Fermi gas vary, as one passed through the intermediate coupling regime.
\par
%%%%%%%%%%%%%%%%%%%%%%%%%%%%%%%%%%%%%%%%%%%%%%%%%%%%%%%%%%%%%%%%%%%%%%%%%%%%%%
\par
\section{Summary}
\par
To summarize, we have discussed normal state properties of a one-component Fermi gas with a $p$-wave interaction. Including $p$-wave pairing fluctuations within the framework of the NSR theory, we calculated the specific heat $C_V$ at constant volume, from the weak-coupling regime to the strong-coupling regime, above the superfluid phase transition temperature $T_{\rm c}$.
\par
At $T_{\rm c}$, we found that $p$-wave pairing fluctuations remarkably enhance the specific heat $C_V$ around $(k_{\rm F}^3v)^{-1}=-7.5$. In the weaker coupling regime, $C_V(T_{\rm c})$ is reduced to the specific heat $C_V^{\rm F}$ in a free Fermi gas. In the stronger coupling side, $C_V(T_{\rm c})$ approaches the value of the specific heat $C_V^{\rm B}$ in an ideal Bose gas, consisting of three kinds of $N/6$ $p$-wave molecules. 
\par
We showed that the enhancement of the specific heat seen at $T_{\rm c}$ soon disappears with increasing the temperature above $T_{\rm c}$, giving a dip structure in the temperature dependence. Using this, we introduced the characteristic temperature $T_{\rm dip}$ as the temperature at which $C_V(T)$ exhibits a dip. In the weak-coupling side ($(k_{\rm F}^3v)^{-1}<0$), $p$-wave pairing fluctuations are strong below $T_{\rm dip}$, causing the anomalous enhancement of $C_V(T\simeq T_{\rm c})$. 
\par
The dip temperature $T_{\rm dip}$ is also obtained in the strong-coupling regime, However, the physical meaning is different from that in the weak-coupling side. In the strong-coupling regime, $C_V(T)$ is well described by the specific heat $C_V^{\rm B}(T)$ of an ideal Bose gas below $T_{\rm dip}$. $C_V(T)$ gradually deviates from $C_V^{\rm B}$ above $T_{\rm dip}$, reflecting the onset of thermal dissociation of molecules into Fermi atoms. 
\par
Using these results, we drew the phase diagram of a $p$-wave interaction Fermi gas in terms of the temperature and the interaction strength. In the normal state above $T_{\rm c}$, this phase diagram has (1) the normal Fermi gas regime, (2) region with strong $p$-wave pairing fluctuations, and (3) molecular Bose gas regime. Strictly speaking, although these regions are not accompanied by any phase transition, this phase diagram would be still useful in considering how a $p$-wave interaction affects normal-state properties of this system, from the temperature dependence of $C_V(T)$. The current experiments in cold Fermi gas physics can only access the normal phase of a $p$-wave interacting Fermi gas. In addition, the observation of the specific heat has recently become possible in this field. Thus, our results would contribute to the study of strong-coupling properties of a $p$-wave interacting Fermi gas within the current experimental technology in this research field.
\par
%%%%%%%%%%%%%%%%%%%%%%%%%%%%%%%%%%%%%%%%%%%%%%%%%%%%%%%%%%%%%%%%%%%%%%%%%%%%%%%
\par
\begin{acknowledgments}
We thank M. Hanai, H. Tajima, T. Yamaguchi, M. Matsumoto, and D. Kagamihara for discussions. This work was supported by KiPAS project in Keio University. DI was supported by Grant-in-aid for Scientific Research from MEXT in Japan (No.JP16K17773). YO was supported by Grant-in-aid for Scientific Research from MEXT and JSPS in Japan (No.JP15H00840, No.JP15K00178, No.JP16K05503). 
\end{acknowledgments}
\par
%%%%%%%%%%%%%%%%%%%%%%%%%%%%%%%%%%%%%%%%%%%%%%%%%%%%%%%%%%%%%%%%%%%%%%%%%%%%%%%
\par


\begin{thebibliography}{99}
\bibitem{note2} In the normal phase, physical quantities are actually insensitive to the detailed momentum dependence of $F_{\rm c}({\bm p})$\cite{Inotani2}, as far as $p_{\rm c}$ is taken to be much larger than the Fermi momentum $k_{\rm F}$.
\bibitem{note} The $x$-axis is chosen to be parallel to an external magnetic field to adjust a $p$-wave Feshbach resonance.
\bibitem{notezz} Here, ``1.01" has been chosen as a typical value to introduce ${\bar T}$, so that this definition, of course, involves ambiguity. However, we believe that it is still useful to identify the region where $C_V(T)\simeq C_V^{\rm B}(T)$.
\end{thebibliography}
\end{document}